\documentclass{jfm} 

\usepackage{amsfonts,amssymb}
\usepackage{amsmath}
\usepackage{latexsym}
\usepackage{graphicx}
\usepackage{color}
\usepackage{natbib}

\usepackage{soul}

\newcommand{\eps}{\epsilon}
\newcommand{\Ups}{\Upsilon}

\title[Near-inertial-wave concentration]{On the concentration of near-inertial waves in anticyclones}
\date{\today}
\author[E. Danioux, J. Vanneste \& O. B\"uhler]{Eric Danioux$^1$, Jacques Vanneste$^1$, Oliver
  B\"uhler$^2$}
\affiliation{$^1$School of Mathematics and Maxwell Institute for
  Mathematical Sciences \\ University of Edinburgh, UK\\
  $^2$Courant Institute of Mathematical Sciences \\ New York
  University, USA\\}

\begin{document}

\maketitle

\begin{abstract}
\noindent
An overlooked conservation law for near-inertial waves
propagating in a steady background flow provides a new perspective on
the concentration of these waves in regions of anticyclonic
vorticity. The conservation law implies that this concentration
is a direct consequence of the decrease in spatial scales experienced by an
initially homogeneous wave field. Scaling arguments and numerical
simulations of a reduced-gravity model of mixed-layer
near-inertial waves confirm this interpretation and elucidate the
influence of the strength of the background flow relative to the dispersion.

\end{abstract}

\hrule

\section{Introduction}
Near-inertial waves (NIWs) are ubiquitous in the ocean. They
contribute strongly to surface mixing, and hence to biological
activity, e.g. \citet{Granata95}.
They also propagate to depth where they eventually
dissipate, thus participating in deep vertical mixing and the global overturning circulation \citep{Ferrari2009}. The propagation of NIWs in heterogeneous flows has
motivated a great deal of work. One of the main conclusions,
emerging from both numerical simulations
\citep{Lee98,Zhai05,Danioux08a} and observational data
\citep{Kunze84,Elipot2010,Joyce2013} is that near-inertial energy
concentrates in 
anticyclones, i.e., in regions of negative relative vorticity in
the Northern hemisphere and positive relative vorticity in the Southern
hemisphere.

Various explanations have been advanced for this
phenomenon. Using a WKB approach, \cite{Kunze85} showed that
refraction by a background flow leads to an `effective' inertial
frequency $f_e$ shifted from the local Coriolis frequency $f$ by
$\zeta/2$, where $\zeta$ is the local relative vorticity:
$f_e=f+\zeta/2$. Because the range of allowed wave frequencies
$\omega$ satisfies $|f_e|<\omega<N$, where $N$ is the
stratification frequency (usually much larger than $|f|$), it is
larger in regions of negative vorticity in the Northern
hemisphere. Therefore, anticyclonic regions are expected to be
more energetic because (i) NIWs propagating poleward can enter
regions of negative vorticity regions, but not regions of
positive vorticity; and (ii) frequencies lower than $f$ present
in the forcing can become resonant \citep{Kunze85}.  This
explanation is subject to caution, however, because the initial
length scales of NIWs, set by atmospheric storms, are typically much larger
than the length scales of geostrophic flows (the ratio of 
tropospheric to  oceanic radii of deformation is about $10$),
thus invalidating the assumption underlying the WKB
approximation.

Alternative explanations were offered by \citet{YBJ} and
\citet{Klein04}. These rely on the NIW model developed by
\citet{YBJ}, hereafter refer to as YBJ model, which exploits the
small frequency spread of NIWs near $f$ and makes no assumption
about their spatial scales.  These explanations rely on strong
assumptions: strong dispersion for \citeauthor{YBJ}'s, short-time
and restrictions on the spectrum of the vorticity field for
\citeauthor{Klein04}'s.  In this note, we revisit the issue and,
taking the YBJ model as a starting point, show that a so-far
overlooked conservation law provides a robust argument for the
concentration of NIWs in anticyclones.  Section \ref{sec:YBJ}
derives the conservation law in the simple case of a
reduced-gravity shallow-water model, the YBJ approximation of
which is obtained in Appendix \ref{sec:derivation}. (The
extension to a continuously stratified fluid is straightforward
when the background flow is assumed barotropic so that NIWs can 
be expanded in vertical modes.)  Section
\ref{sec:concentration} demonstrates analytically and numerically
how NIW concentration in anticyclones stems from the conservation
law. Some conclusions and perspectives are offered in section
\ref{sec:conclusion}.

\section{Model and conservation laws}\label{sec:YBJ}
\subsection{YBJ model and analogy with the Schr\"odinger equation}
We study the propagation of NIWs in a steady geostrophic flow
with a reduced-gravity shallow-water model (e.g.,
\citealt{Cushman}). This slab-model can be thought of as
representing the dynamics of NIWs confined in a mixed layer
capping an abyssal layer where the only motion is the imposed
geostrophic flow, assumed to be identical in both layers.  For
NIWs, the horizontal velocity $(u,v)$ can be written in terms of
a complex amplitude $M$ according to $u+iv= M e^{-ift}$. The slow
(compared to $f^{-1}$) time evolution of $M$ is governed by the
YBJ equation
\begin{equation}\label{eq:YBJ}
\partial_t M + J(\psi,M) - i\frac{h}{2}\Delta M + i\frac{\Delta \psi}{2} M=0,
\end{equation}
where $\psi$ and $\Delta \psi$ are the streamfunction and
vorticity of the steady 
geostrophic flow, $J$ is the horizontal
Jacobian, and $h=g'H/f$ is a dispersion parameter, with $g'$ and $H$ the reduced gravity
and average depth of the mixed layer. We assume that $h>0$, as
in 
the Northern hemisphere where $f>0$. 
The respective terms quantify the effects due to advection,
dispersion, and refraction.  A concise derivation of
(\ref{eq:YBJ}) is given in Appendix \ref{sec:derivation}.

Two simple facts help in understanding the dynamics of
(\ref{eq:YBJ}).  First, for any constant $\alpha >0$ the YBJ
equation is invariant under the scaling transformation
$\psi\mapsto\alpha\psi$, $h\mapsto\alpha h$, $t\mapsto t/\alpha$.  This makes
obvious that the intrinsic dynamics of (\ref{eq:YBJ}) depends
only on the single non-dimensional parameter $h/\Psi$, say, where
$\Psi$ is the amplitude scale of $\psi$. 

 Second, without the advection
term $J(\psi,M)$ the YBJ equation is identical to the
Schr\"odinger equation that governs the complex wave function
$\phi(x,y,t)$ for  a single particle with unit mass 
and external potential $V(x,y)$:
\begin{equation}
  \label{eq:1}
  \partial_t \phi - i\frac{\hbar }{2}\Delta \phi +
  i\frac{V}{\hbar}\phi =0.
\end{equation}
Here $\hbar$ is Planck's constant divided by $2\pi$.  Comparing
(\ref{eq:YBJ}) and (\ref{eq:1}) and identifying $\hbar$ with
$h$ shows that the effective potential in (\ref{eq:YBJ}) is 
\begin{equation}
  \label{eq:2}
  V = h\frac{\Delta \psi}{2}.
\end{equation}
Clearly, regions of higher $V$ repel the particle whereas
regions of lower $V$ attract it.  Hence, if the advection
term is negligible, then the mathematical analogy between the
particle probability density $|\phi|^2$ and the inertial wave
kinetic energy density $|M|^2$ readily implies that cyclones
repel inertial waves whilst anticyclones attract them
\citep{Balmforth98}.  
For the special case of an axisymmetric anticylonic vortex, this is confirmed by the existence of axisymmetric trapped modes \citep{Llewellyn99}.

Of course, if the advection term cannot be
neglected then the simple analogy with the Schr\"odinger
equation breaks down.  
Hence, the main task is to understand how the advection term
alters  the basic Schr\"odinger dynamics as a function of $h/\Psi$.   
For this it becomes crucial to study the full
set of conservation laws associated with (\ref{eq:YBJ}), as we
shall do now.

\subsection{Conservation laws}
\label{sec:consv}

Multiplying (\ref{eq:YBJ}) by $M^*$ and adding its complex
conjugate gives
\begin{equation}\label{eq:KE}
\partial_t \frac{1}{2}|M|^2+J(\psi,\frac{1}{2}|M|^2)+\boldsymbol{\nabla}\cdot\mathbf{F}=0,
\end{equation}
where $\mathbf{F}=ih(M\boldsymbol{\nabla}M^*-M^*\boldsymbol{\nabla}M)/4$.
In a finite domain with suitable boundary conditions (periodicity or $M=0$),
integrating (\ref{eq:KE}) gives the conservation of NIW-kinetic
energy, as derived by \cite{YBJ},
\begin{equation}\label{eq:KE_conservation}
\frac{d}{dt}\iint \frac{1}{2}|M|^2dxdy=0.
\end{equation}
There is another conservation law associated with (\ref{eq:YBJ})
and not mentioned in \citet{YBJ}. It is derived by forming the
combination $M_t^*(\ref{eq:YBJ})-M_t(\ref{eq:YBJ})^*$ and
integrating the result over the domain. Using properties of the
Jacobian and integrating by parts, this gives
\begin{equation}\label{eq:PE_conservation}
\frac{d}{dt} (I_1+I_2+I_3)=0,
\end{equation}
where 
$$
I_1=\iint i h\psi J(M^*,M)dxdy, \ I_2=\iint \frac{h^2}{2}|\boldsymbol{\nabla}M|^2dxdy, \  I_3=\iint h\frac{\Delta \psi}{2}|M|^2dxdy.
$$
The terms $I_1$, $I_2$ and $I_3$ stem directly from the
advection, dispersion and refraction terms in the YBJ equation.  
The overall factors of $h$ are included for two practical
reasons: to keep the values of the invariants comparable when $h$
is varied, and to highlight the appearance of the effective
potential $V$ from (\ref{eq:2}) in $I_3$.  
The consequences of the new conservation law
(\ref{eq:PE_conservation}) are discussed in more detail in the
next section.  In the absence of a geostrophic flow,
$I_1, I_3$ are identically zero, and $I_2$ can be recognised as
the 
scaled %
NIW potential energy averaged over the fast time scale.
This is not unexpected, as explained in
Appendix \ref{sec:energy}. In the presence of a steady flow, $I_1+I_2+I_3$ can be interpreted as an energy in that its conservation is associated with the time invariance of (\ref{eq:YBJ}). Unlike (\ref{eq:KE_conservation}),  $I_1+I_2+I_3$ is not conserved for arbitrary time-dependent flows. However, when flow and NIWs evolve in a dynamically consistent manner, an analogous conservation  law holds that accounts for energy transfers between flow and NIWs \citep{Xie2015}.

We remark that, in a steady geostrophic flow, differentiating
(\ref{eq:YBJ}) with respect to time shows that $M_t$ satisfies
the same equation as $M$, and hence the same conservation
laws. In particular,
\begin{equation}\label{eq:KEt_conservation}
\frac{d}{dt} \iint \frac{1}{2}|M_t|^2dxdy=0
\end{equation}
means that the root-mean-square magnitude of
$M_t$ is constant.

\section{NIW-concentration in anticyclones}\label{sec:concentration}

Non-dimensionalizing (\ref{eq:YBJ}) using $x=Lx'$,
$\psi=\Psi\psi'$, 
$h=\Psi h'$, and $t=(L^2/\Psi)t'$, with $L$ the typical length
scale of the geostrophic flow, 
gives an identical equation for the primed variables.  Again,
this
makes obvious that $h' = h / \Psi$ is the only relevant
parameter; in \S\ref{sec:num} we conduct
simulations with different values of $h/\Psi$.  This parameter is the
reduced-gravity shallow-water equivalent to the parameter
$\Ups=\Psi/h=1/h'$ used in \cite{YBJ} and \cite{Balmforth98}, on which
they base their `strong dispersion' ($h/\Psi\gg 1$) and `strong
trapping' (or `strong advection', $h/\Psi\ll 1$) approximations.  In the ocean, $h/\Psi$ is highly
variable because of varying kinetic-energy levels and
stratification. For instance, typical values for the North
Atlantic might be $f=10^{-4}\, \text{s}^{-1}$, $g'=2\cdot10^{-3}\,
\text{m}\,\text{s}^{-2}$, $L=50$ km; taking $H$ and $U$ in the
ranges $H \in [50,100]$ m, $U\in[0.005,0.1] \,
\text{m}\,\text{s}^{-1}$ gives $h/\Psi\in [0.2, 8]$. 

We 
will consider a specific initial-value problem in which
$M(x,y,0)=1$.  Without loss of generality this represents
an eastward NIW-momentum deposition by a storm: because storm scales are
typically much larger than ocean eddy scales, a homogeneous
initial condition for $M$ is appropriate. 

The conservation law (\ref{eq:PE_conservation}) involves $I_3$,
which is proportional to the covariance between $|M|^2$ and
$\Delta\psi$; it is therefore relevant to the concentration of
NIWs in anticyclones, which corresponds to $I_3<0$. At $t=0$,
$I_1=I_2=I_3=0$ (assuming no net vorticity -- a given with
periodic boundary conditions).  The development of spatial
heterogeneities in the $M$-field must lead to an increase in the
positive definite $I_2$, which is then compensated by $I_1 +
I_3<0$.  Of course, if $I_1$ is negligible then $I_3$ must take
negative values, but in the
general case it is less clear whether $I_3$ behaves in this way.
%
%
%
We next provide an asymptotic argument that $I_3$ becomes
negative for all values $h/\Psi$ at short times.  This is
followed by a long-time scaling argument for (\ref{eq:YBJ}) as a function of
$h/\Psi$, which predicts that  $I_3<0$ for large
and intermediate values of $h/\Psi$ but not for small values of
this parameter.  These predictions are then
checked against numerical simulations in \S~\ref{sec:num}.

\subsection{Short-time solution}\label{sec:asymptotics}

With homogeneous initial conditions, the first physical effect on
NIWs propagation is due to refraction, and the short-time
behaviour is $M(x,y,t)=\exp(-it\Delta\psi(x,y)/2)\equiv
\bar{M}(x,y,t)$ \citep{Danioux08a}. A cautious definition for
short time here is $t\ll t_s=1/\max\{\Psi/(2L^2),h/(2L^2)\}$. For
such times, the solution can be sought as the expansion
$M=\bar{M}+M'$, with $M'(x,y,t=0)=0$ and $|M'|\ll1$.  Introducing this
into (\ref{eq:YBJ}) gives
\begin{equation}\label{eq:M'}
\partial_t M' + J(\psi,M') - i\frac{h}{2}\Delta M' +
i\frac{\Delta \psi}{2} M'=-J(\psi,\bar{M})+i\frac{h}{2}\Delta
\bar{M}. 
\end{equation}
Because of the form of $\bar{M}$, the short-time behaviour of the
right-hand side of (\ref{eq:M'}) behaves as $O(t)$, hence forcing
$M'(t)=O(t^2)$ for small $t$.  Keeping this in mind, we now
compare the relative size of terms $I_1$ and $I_2$ in
(\ref{eq:PE_conservation}). Firstly, because
$J(\bar{M}^*,\bar{M})=0$, $I_1$ is dominated by terms of the form
$ih\psi J(\bar{M}^*,M')$, resulting in a $O(t^3)$
dependence. Secondly, $I_2$ is dominated by
$h^2|\boldsymbol{\nabla}\bar{M}|^2/2$, yielding a $O(t^2)$
dependence. Hence, for times short enough,
$I_2 \gg I_1$.  NIW-energy concentration in regions of negative
vorticity can also be deduced from (\ref{eq:KE}). Injecting the
short-time solution $\bar{M}(x,y,t)$ into $\mathbf{F}$, one finds
that the amplitude of $M$ obeys
\begin{equation}\label{eq:KE_short_time}
\partial_t \frac{1}{2} |M|^2\simeq\frac{ht}{4}\Delta^2\psi
\end{equation}
at short times, consistent with \citet{Klein04}'s short-time
solution. Because the vorticity field and its Laplacian are
anticorrelated, (\ref{eq:KE_short_time}) gives an increase of
NIW-energy in anticyclonic regions 
for all values of $h/\Psi$.

\subsection{Long-time scaling arguments }
\label{sec:long-time-scaling}

For long times the spatial scales of $M$ need not be equal to
those of $\psi$ anymore and hence the dominant balance between
the various terms in (\ref{eq:YBJ}) may shift accordingly.
However, the conservation law (\ref{eq:KEt_conservation}) implies 
\begin{equation}
  \label{eq:4}
  \iint \frac{1}{2}|M_t|^2dxdy=\iint
\frac{1}{2}|M_t(t=0)|^2dxdy=\iint (\Delta\psi/2)^2dxdy
\end{equation}
and together with 
(\ref{eq:KE_conservation}) this means that for all times the
root-mean-square magnitudes of $M$ and $M_t$ scale with unity and
$\Psi/L^2$, respectively.  This allows a simple scaling analysis of the four
terms in (\ref{eq:YBJ}), which after rearranging yields
\begin{equation}
  \label{eq:3}
  1,\quad  \quad \frac{L}{l}, \quad  \quad
  \frac{h}{\Psi}\frac{L^2}{l^2},\quad  \quad 1.
\end{equation}
Here $l$ is the long-time spatial scale of $M$ such that $\nabla
M=O(1/l)$.  We use (\ref{eq:3}) to determine how $l/L$ may depend on
$h/\Psi$.  First, in the `strong dispersion' regime $h/\Psi\gg1$
the only possible balance in (\ref{eq:3}) is $l/L
=\sqrt{h/\Psi}$, which balances dispersion and refraction 
whilst advection is negligible.  As expected, this
reduces the YBJ dynamics to that of the Schr\"odinger equation,
so $I_1$ is negligible and $I_3\approx - I_2$.  In
this regime the spatial scale of $M$ is larger than that of
the background flow $\psi$ by a factor of $\sqrt{h/\Psi}$.

Second, in the opposite regime $h/\Psi\ll1$ one scaling
possibility is $l=L$, which balances advection and refraction
whilst dispersion becomes negligible.  However, from
(\ref{eq:YBJ}) this would correspond to an advective dynamics
along streamlines in which $|M|$ is conserved whilst its phase
continues to change by the refraction as in the short-term
solution derived in \S~\ref{sec:asymptotics}.  This would
inevitably lead to the generation of evermore smaller spatial
scales in $M$ and hence defeat the assumption $l=L$.  We must
therefore look at the alternative, a balance between
the advective and the dispersive terms based on $l/L=h/\Psi$.
The advective and dispersive terms then provide a new
leading-order dominant balance of size $\Psi/h\gg1$ to the
long-term evolution of the YBJ equation.  Consequently, in this
scenario the refraction is weak and $I_3$ becomes negligible and
so $I_1\approx - I_2$.  Moreover, the spatial scale of $M$ is
much smaller than that of $\psi$, by a factor of $h/\Psi$.

Note that this small horizontal scale may invalidate the
assumption of small $h/(fl^2)$ that underpins the near-inertial
YBJ model (see Appendix \ref{sec:derivation}). Using $l = h
L/\Psi$, we rewrite this assumption as $\Psi/(fL^2)\ll h/\Psi$ in
terms of the Rossby number on the left-hand side. Hence for small
but fixed $h/\Psi$ the near-inertial approximation holds provided
the Rossby number is small enough.

Finally, in the intermediate regime $h/\Psi=1$ the 
scaling $l/L=1$ makes all terms equally important.  It stands to reason
that in this intermediate regime $I_3$ will take moderate
negative values, consistent with a monotonic transition between
its value $I_3\approx 0$ for $h/\Psi\ll1$ and its negative value
$I_3\approx -I_2$ for $h/\Psi\gg1$.  However, only in this
intermediate regime does the length scale of $M$ equal that of
$\psi$, which arguably is the best situation for effective
concentration of NIW energy in anticyclones.  This suggests that
the intermediate regime might be the most effective for this
purpose.  Of course, this simple scaling analysis can only
provide a heuristic guide to the full NIW energy dynamics, not
least because the spatial scales of $|M|^2$ are not related in a
trivial way to those of $M$.  Still, we will see that the present
scaling arguments do indeed provide a useful guide for
understanding the numerical simulations.

\subsection{Numerical simulations}\label{sec:num}

\begin{figure}
\begin{center}
\begin{tabular}{cc}
\includegraphics[scale=.25]{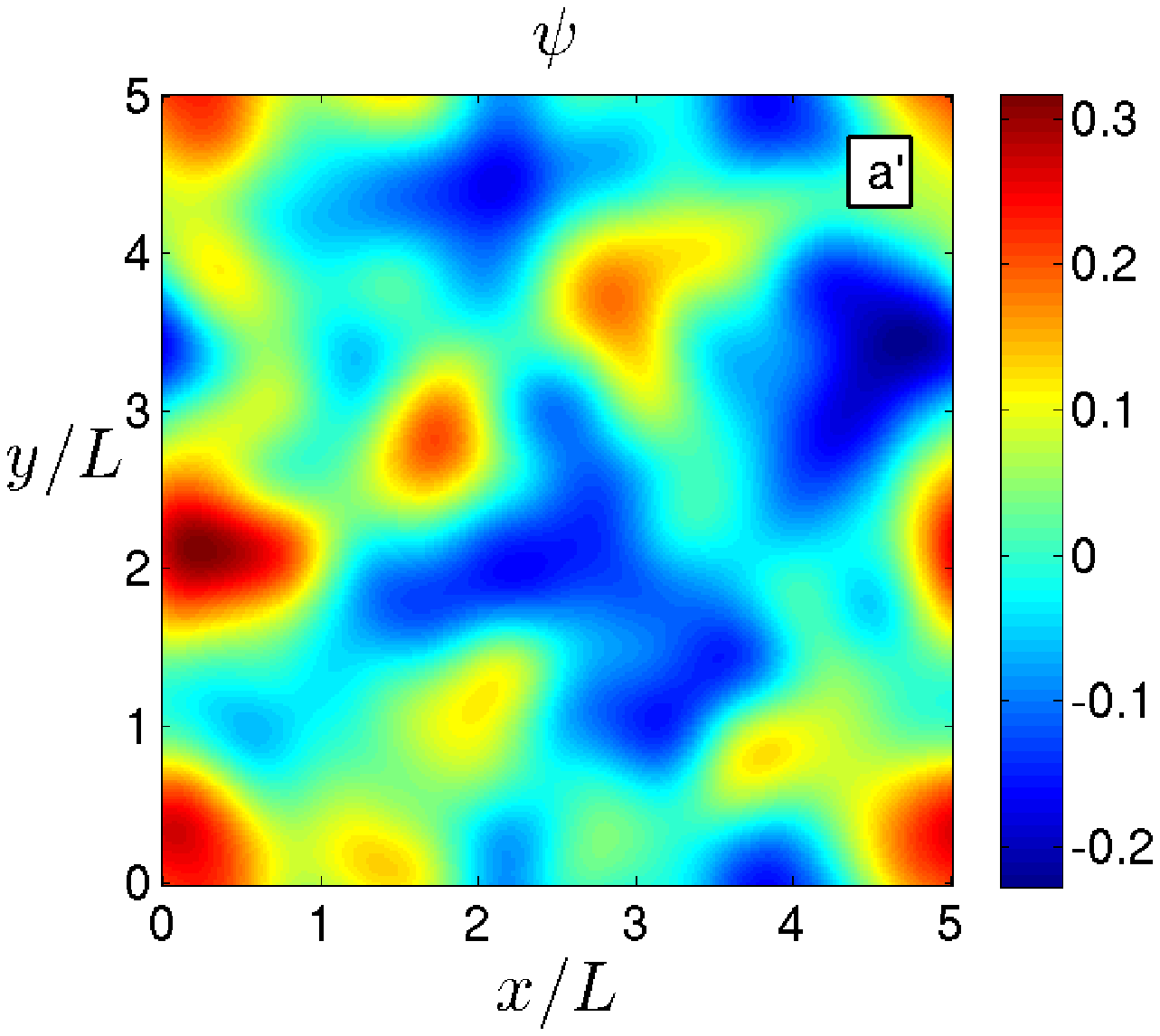}&
\includegraphics[scale=.25]{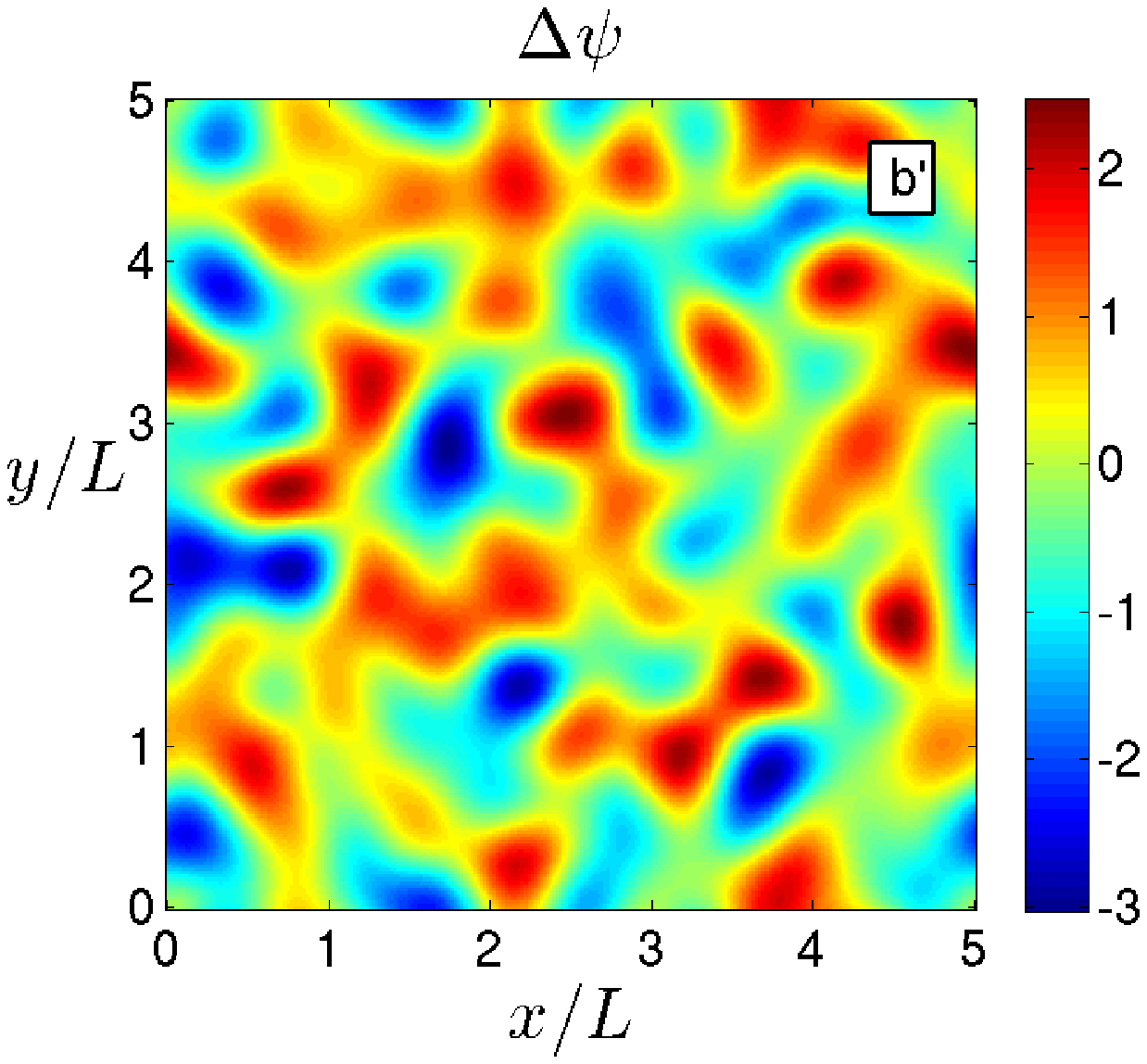}
\end{tabular}
\end{center}
\begin{center}
\begin{tabular}{ccc}
\includegraphics[width=.3\textwidth]{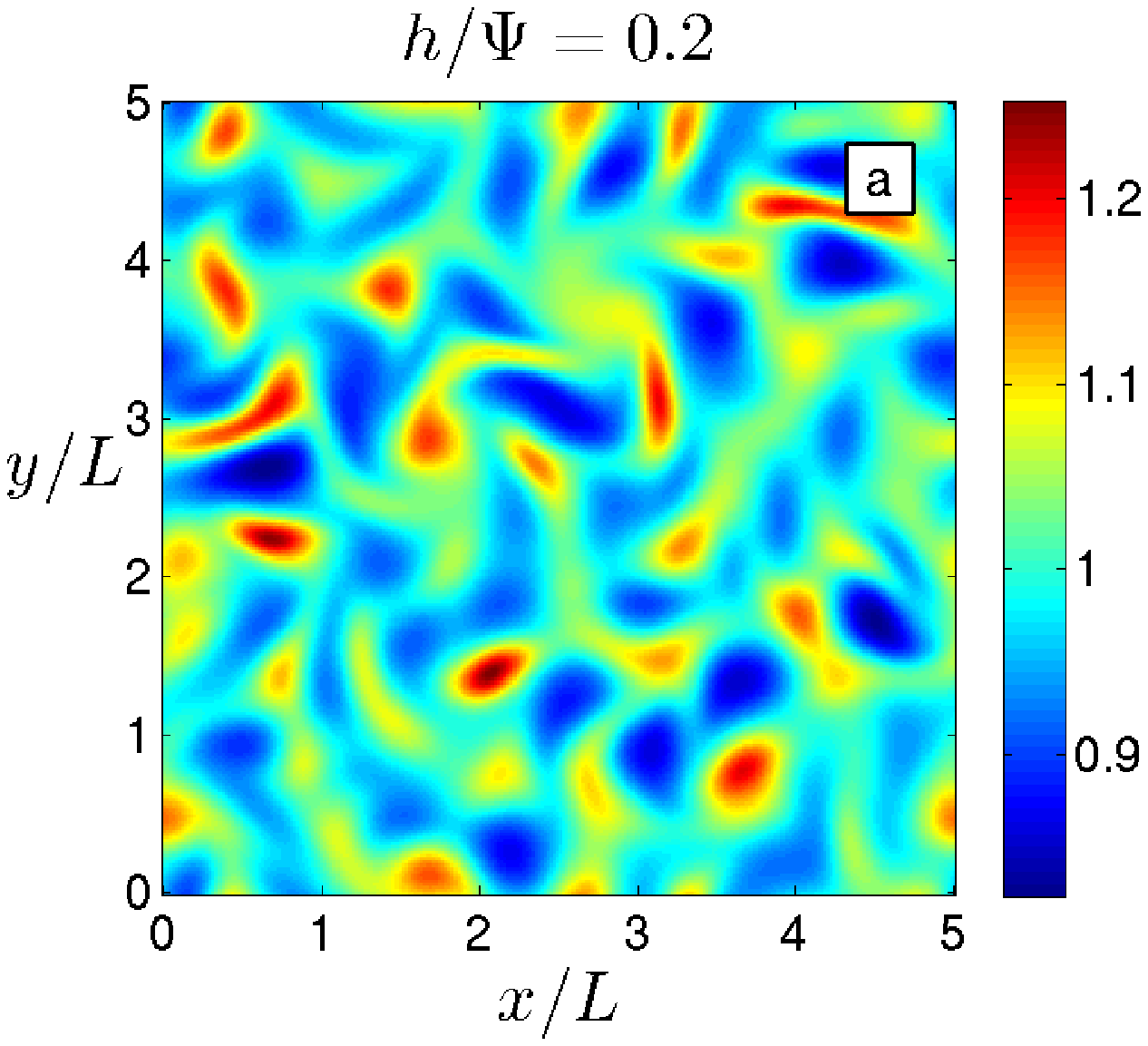}&
\includegraphics[width=.3\textwidth]{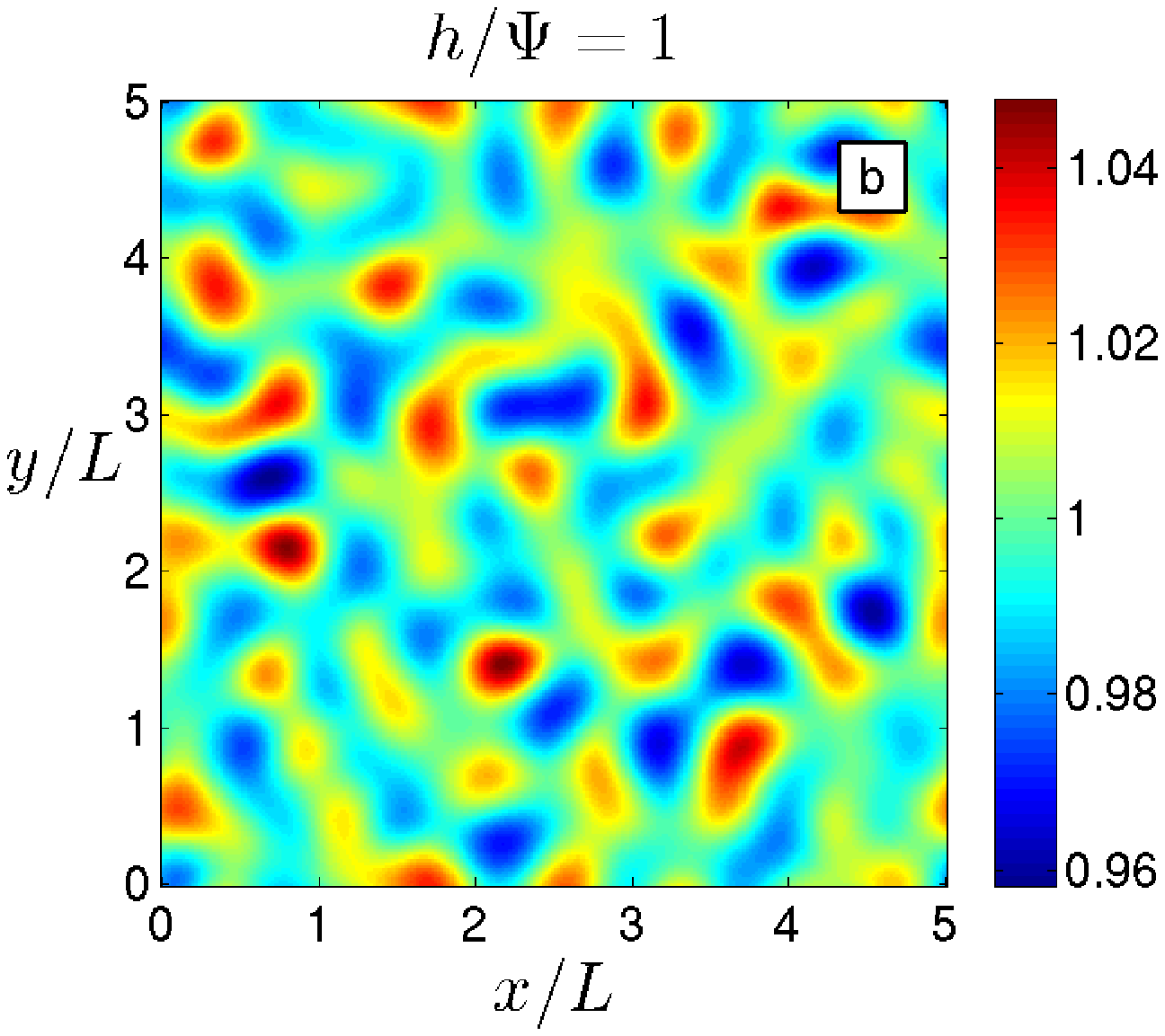}&
\includegraphics[width=.3\textwidth]{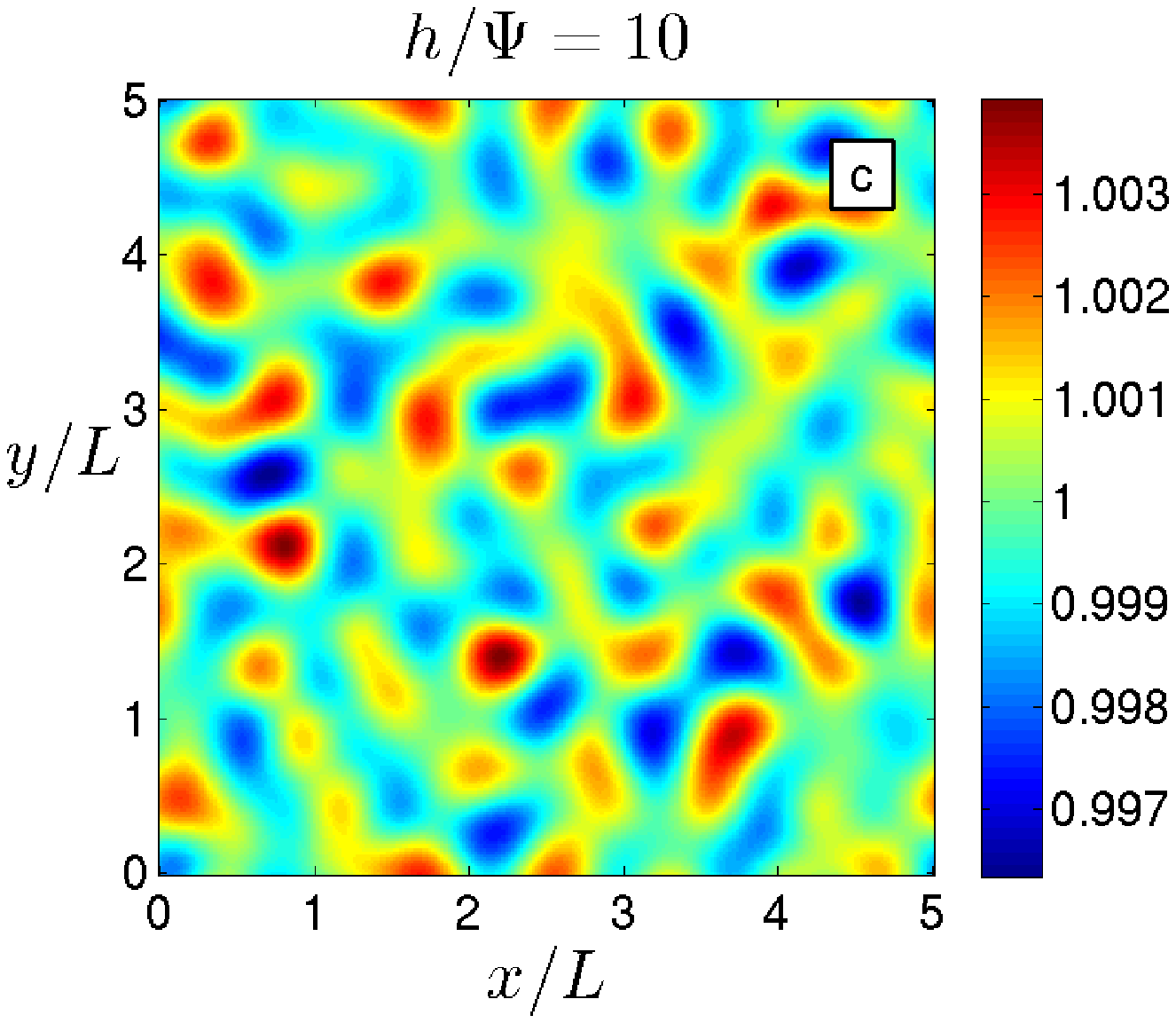}\\
\includegraphics[width=.3\textwidth]{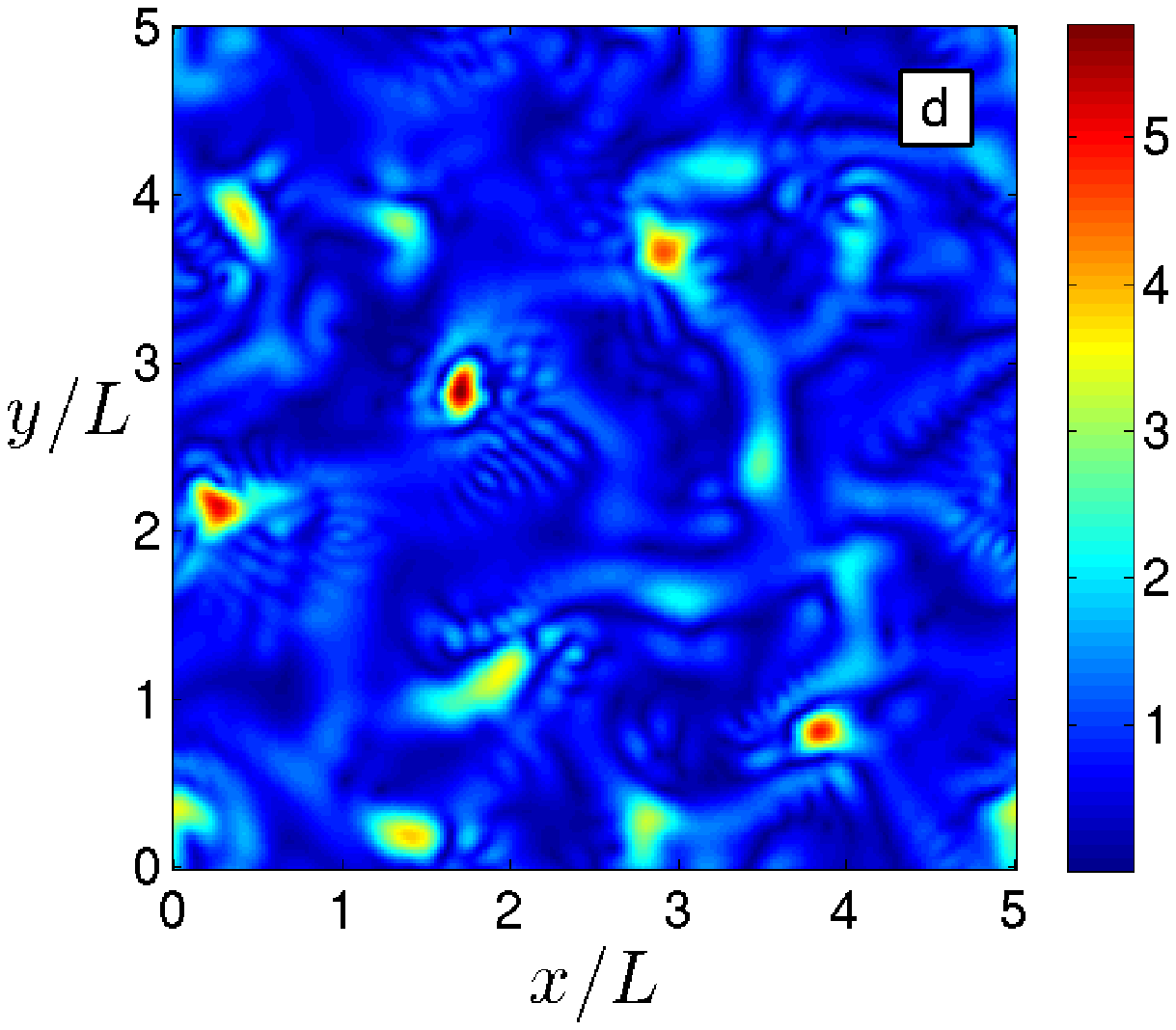}&
\includegraphics[width=.3\textwidth]{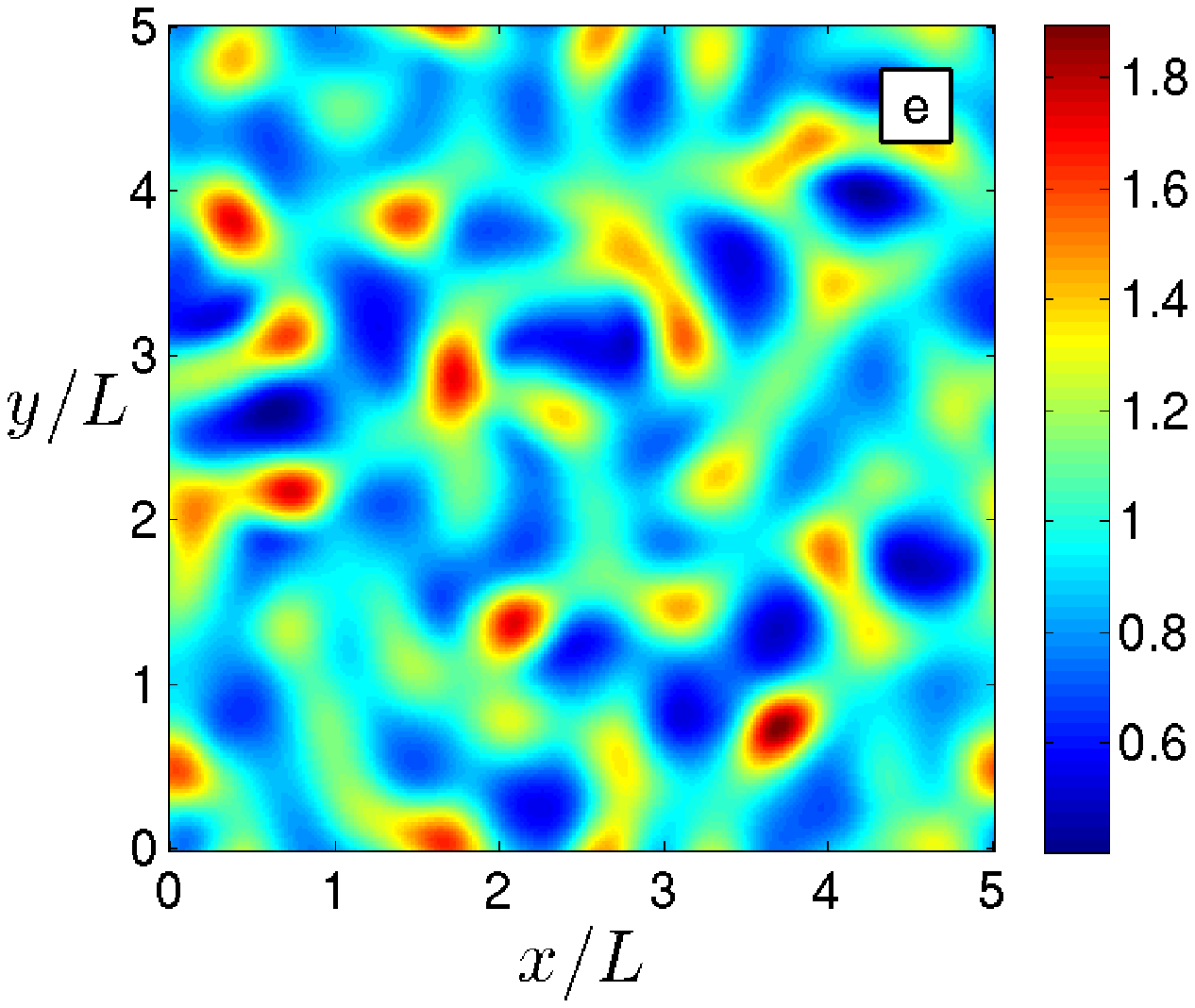}&
\includegraphics[width=.3\textwidth]{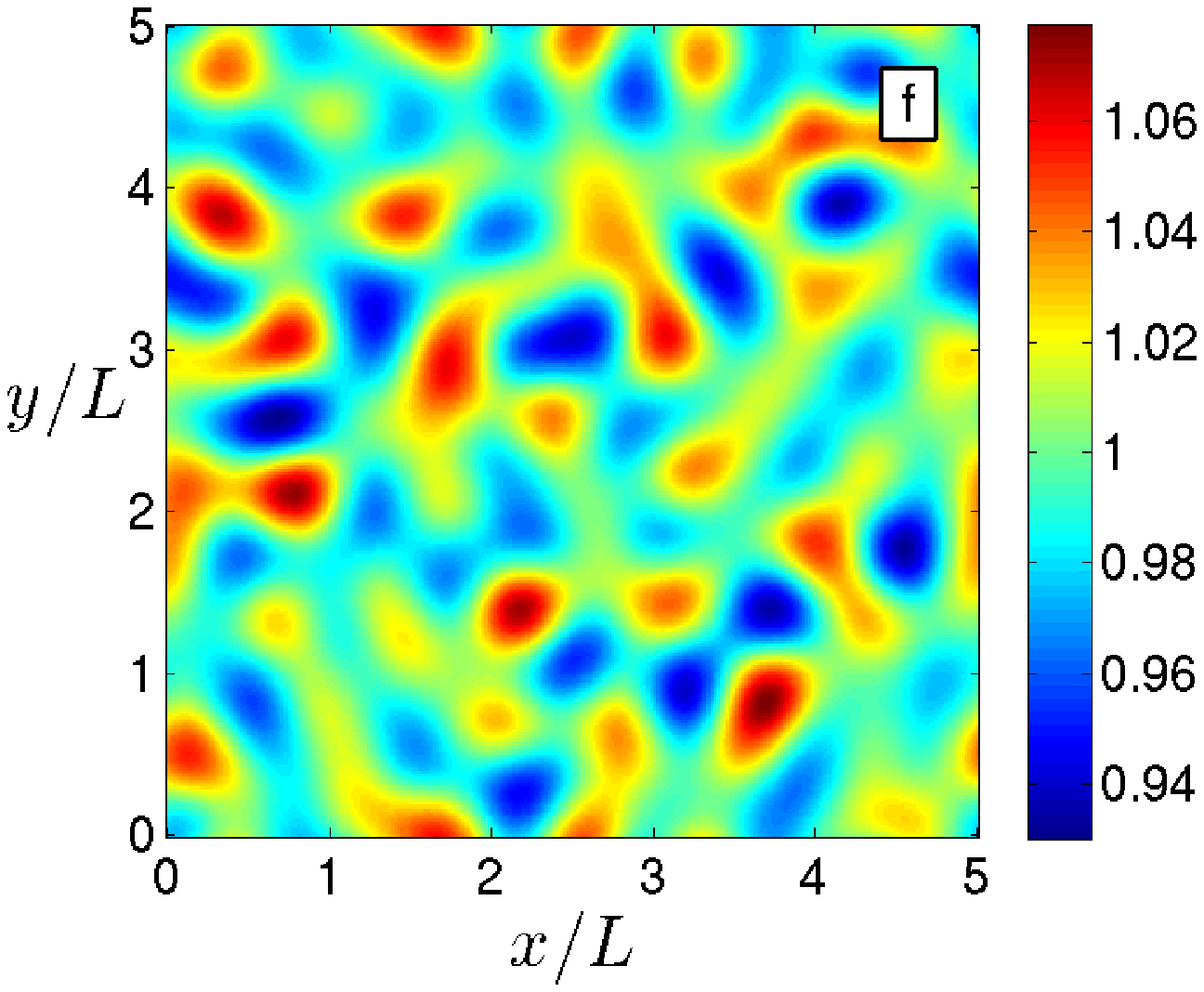}\\
\includegraphics[width=.3\textwidth]{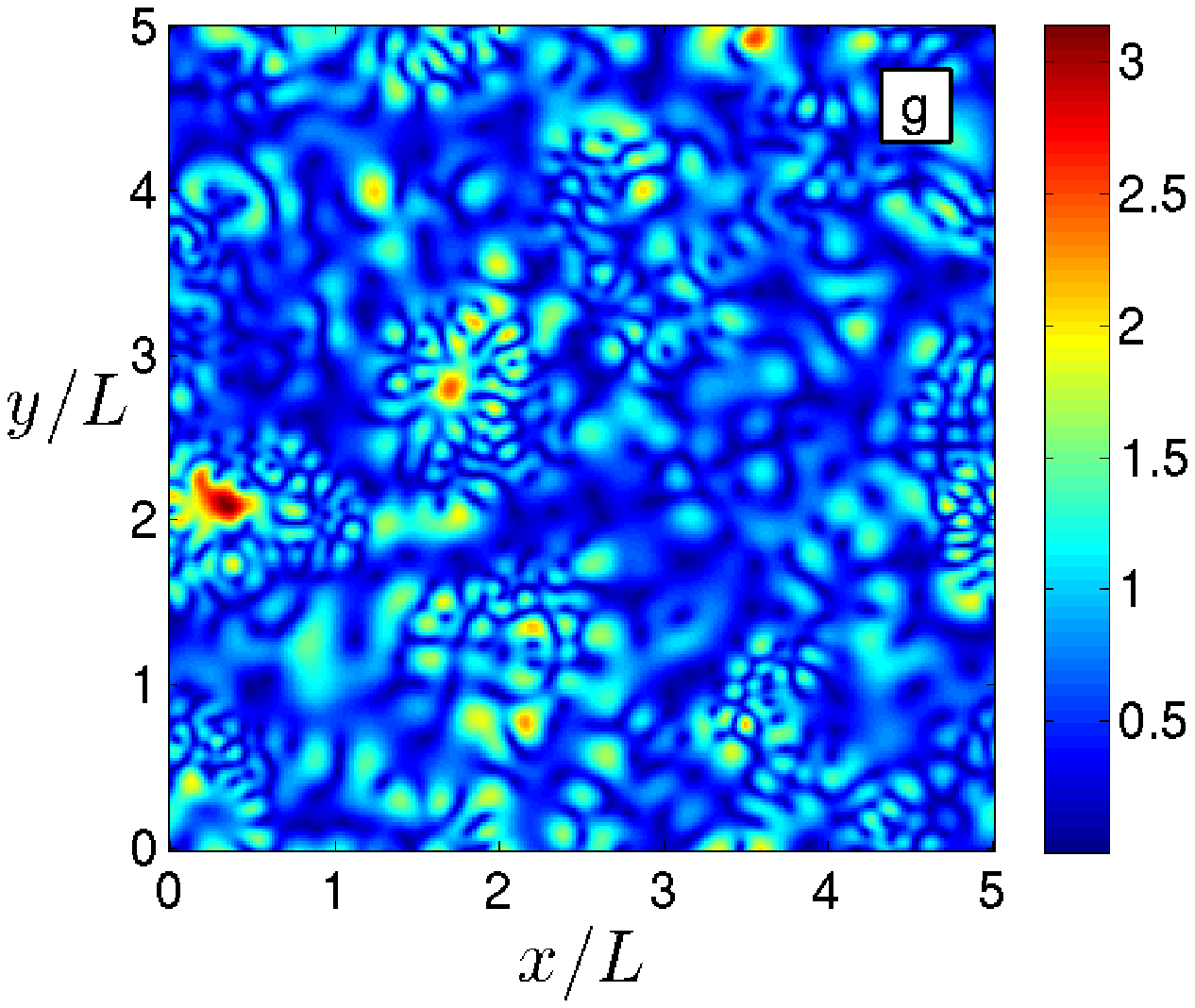}&
\includegraphics[width=.3\textwidth]{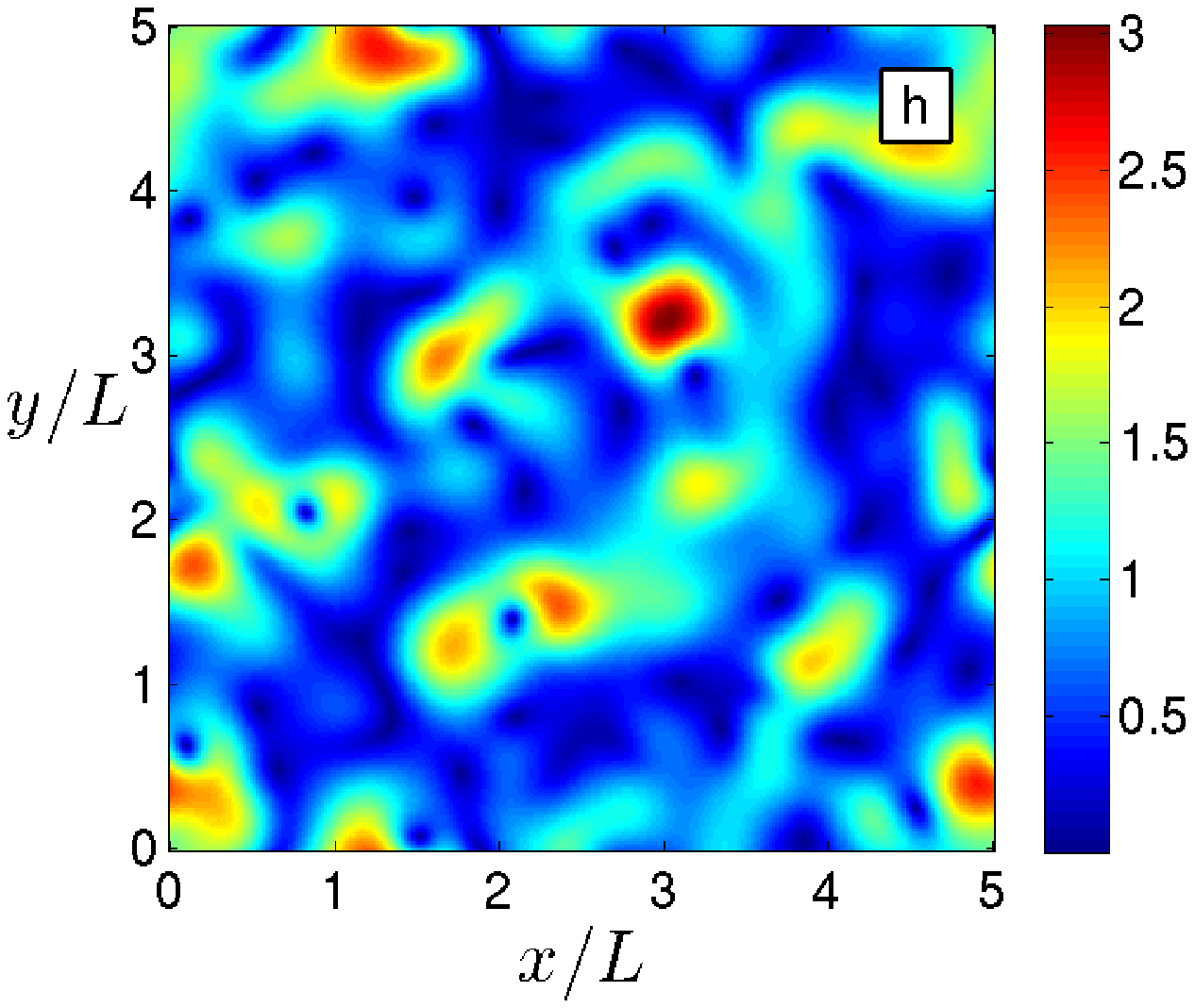}&
\includegraphics[width=.3\textwidth]{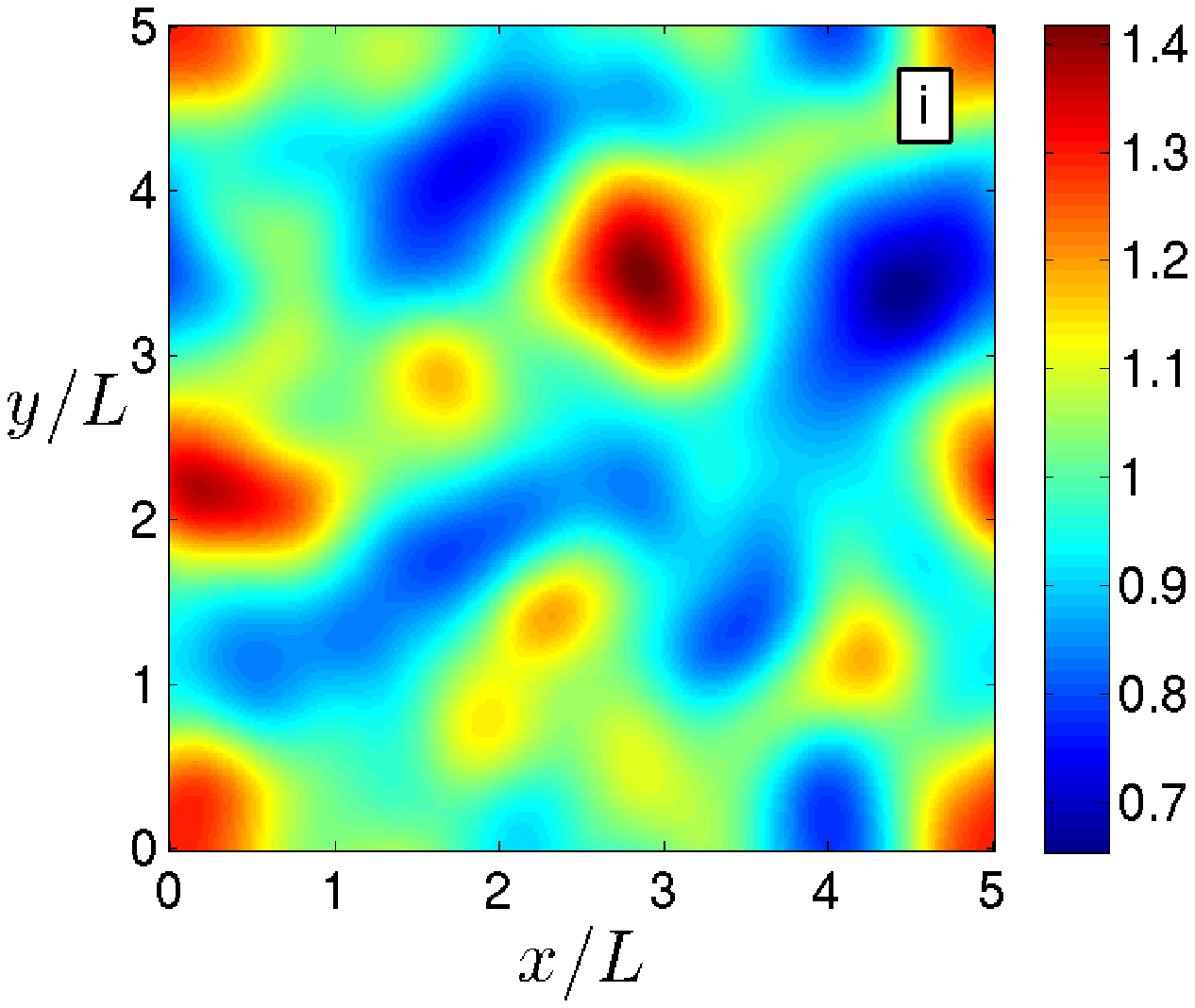}
\end{tabular}
\caption{Streamfunction (a') and associated normalized vorticity field (b') of the background flow, and evolution of the NIW-amplitude $|M|$ for $h/\Psi=0.2$ (left
  column), $h/\Psi=1$ (middle column) and $h/\Psi=10$ (right
  column). Panels (a), (b), (c) correspond to very short
  times after the start of the simulations ($1/80^{\text{th}}$ of
  the length of the simulations), panels (d), (e), (f)
  to short times ($1/16^{\text{th}}$ of the length of
  the simulations) and bottom panels  to the end of the
  simulations.}
\label{fig:vort_M_field}
\end{center}
\end{figure}

Eq.\ (\ref{eq:YBJ}) is solved
numerically on a doubly periodic $256\times256$ grid using a
pseudo-spectral time-split Euler scheme. A weak biharmonic
dissipation is added for numerical stability. The streamfunction
$\psi$ is taken as a %
single realization of a homogeneous isotropic 
Gaussian random process, with Gaussian correlation function and a
correlation length $L$ of about $1/5$ of the domain size.  The chosen vorticity
field is shown at the top of figure
\ref{fig:vort_M_field}. Because its correlation scale is much
smaller than the size of the domain, the results presented here are generic.  We run
three simulations with the same background flow but different
values of $h$ such that $h/\Psi = 0.2$, $1$ and $10$,
representative of the strong advection, intermediate and strong
dispersion regimes. 
Simulations are stopped when the various
terms in (\ref{eq:PE_conservation}) no longer evolve
significantly,
which happens around $t \simeq 0.3L^2/h$ (figure \ref{fig:vort_M_field}). For the
intermediate case, using the values given at the beginning of
this section, we find a time-scale between $4$ and $8$ days,
which is realistic.
 
Snapshots of $|M|$ are shown in figure \ref{fig:vort_M_field} for
very short, short and long times for the three values of
$h/\Psi$. At very short times, for all values of $h$, $|M|$ is
clearly anti-correlated with the vorticity field (or correlated
with its Laplacian), as predicted by (\ref{eq:KE_short_time}).
As time increases, the variance of $|M|$ increases, substantially
for small $h/\Psi$ but much less for large $h/\Psi$. This is
consistent with the scaling $|M-1|=O(\Psi/h)$ that holds for
short time and arbitrary $h/\Psi$ (as follows from integrating
(\ref{eq:KE_short_time}) for $t \lesssim L^2/h$) and for all time
and large $h/\Psi$ (as follows from a perturbative treatment of
(\ref{eq:YBJ})).  At the end of the simulation, $M$ has larger
scales than at the very first instants for $h/\Psi=10$, in
accordance with the scaling arguments in  
\S~\ref{sec:long-time-scaling} and also with \citet{Klein04}'s
`truncated Laplacian' solution. The spectrum of $M$ does not
change much afterwards (not shown), in contrast with
\citet{Klein04}'s theory which predicts a continued cascade
towards large scales.  
As expected,  the evolution is very different for $h/\Psi=0.2$. The
short-time solution (figure \ref{fig:vort_M_field}d) is
consistent with a passive-scalar type scenario: after the initial
generation of spatial scales in $|M|$ (figure
\ref{fig:vort_M_field}a), the $|M|$-field is stretched and
folded, while its amplitude grows due to dispersion. Later on
(figure \ref{fig:vort_M_field}g), NIW-scales much smaller than
the flow scales develop, consistent with
\S\ref{sec:long-time-scaling}. 

\begin{figure}
\begin{center}
\begin{tabular}{ccc}
\includegraphics[width=.3\textwidth]{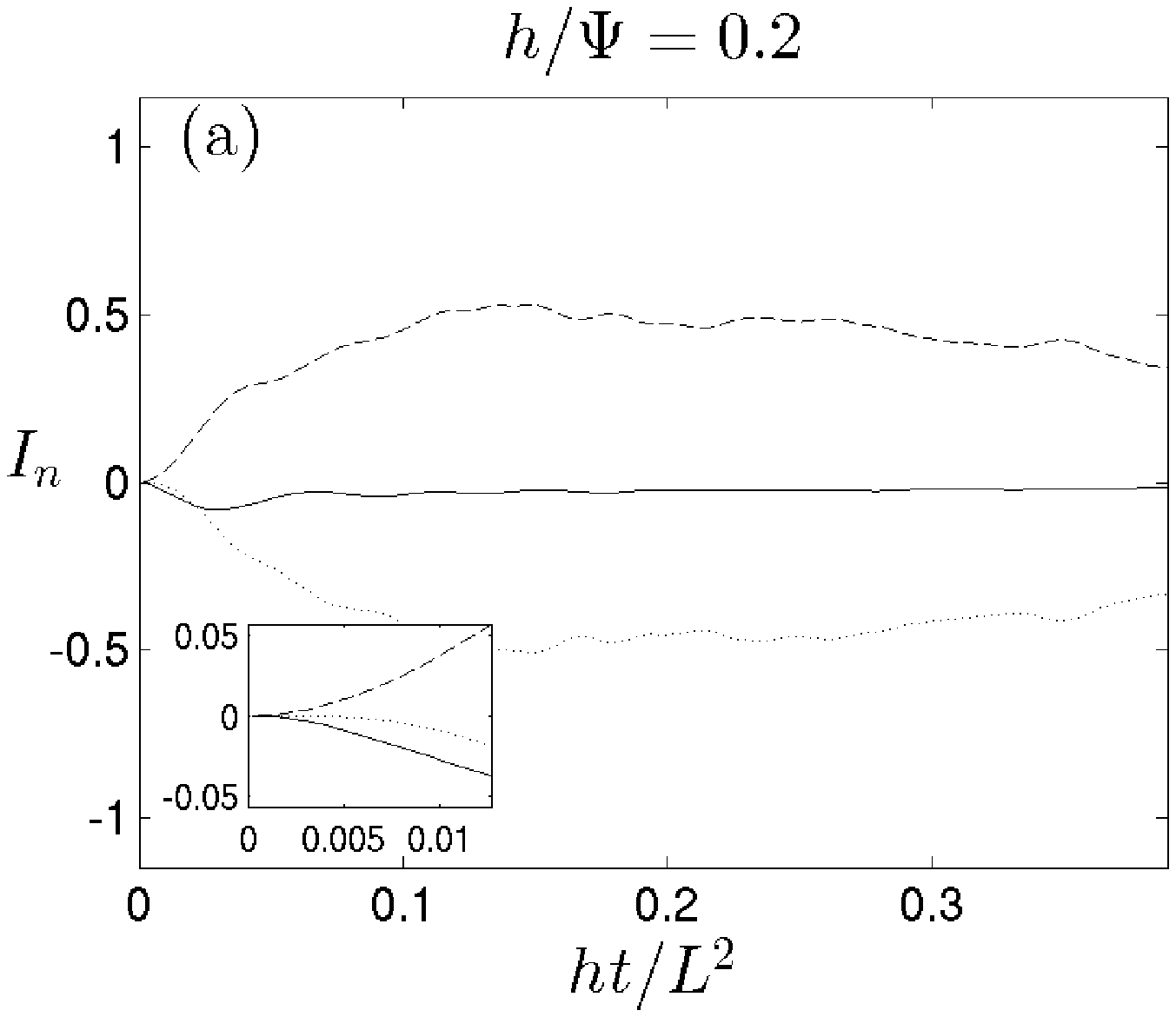}&
\includegraphics[width=.3\textwidth]{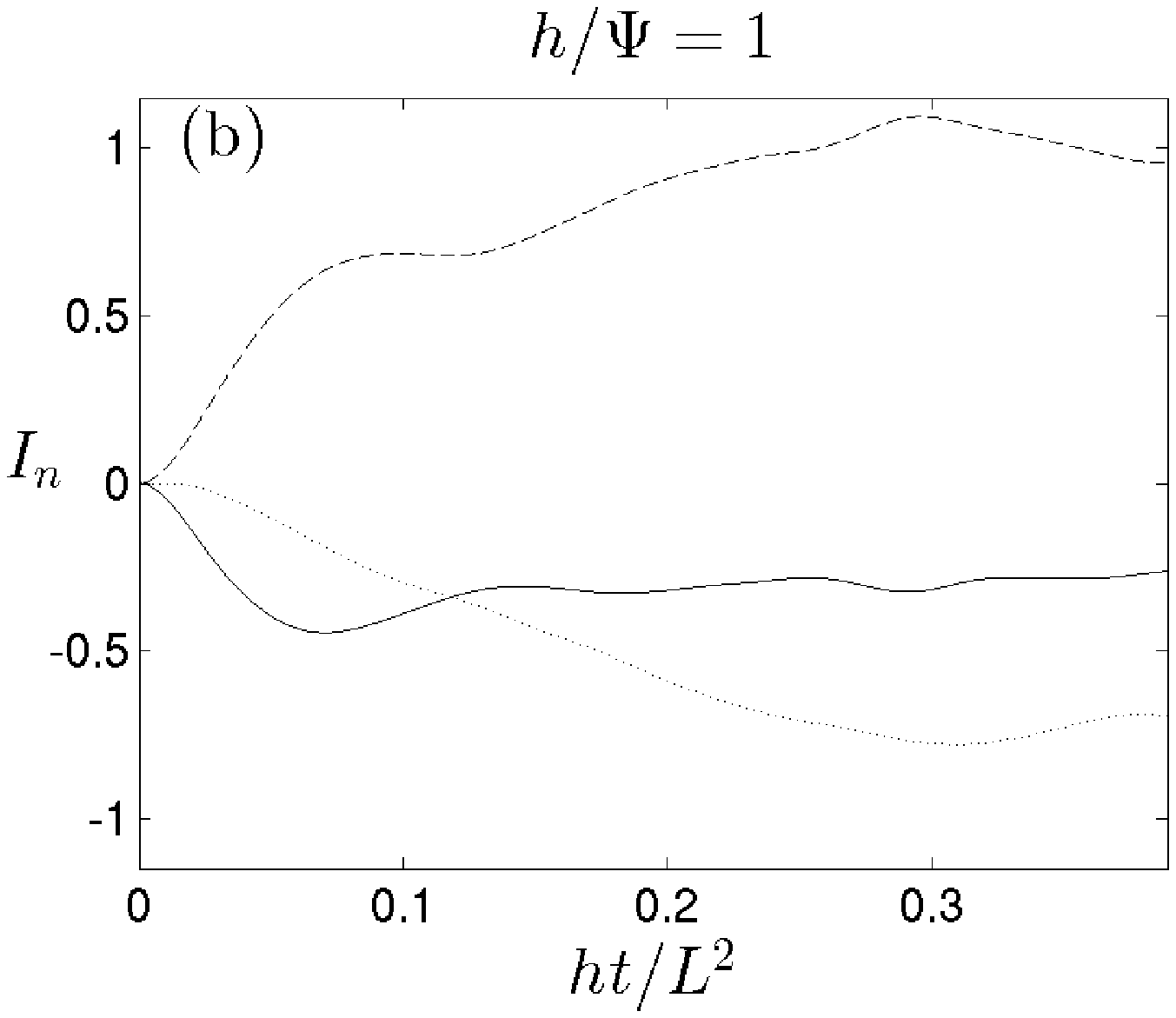}&
\includegraphics[width=.3\textwidth]{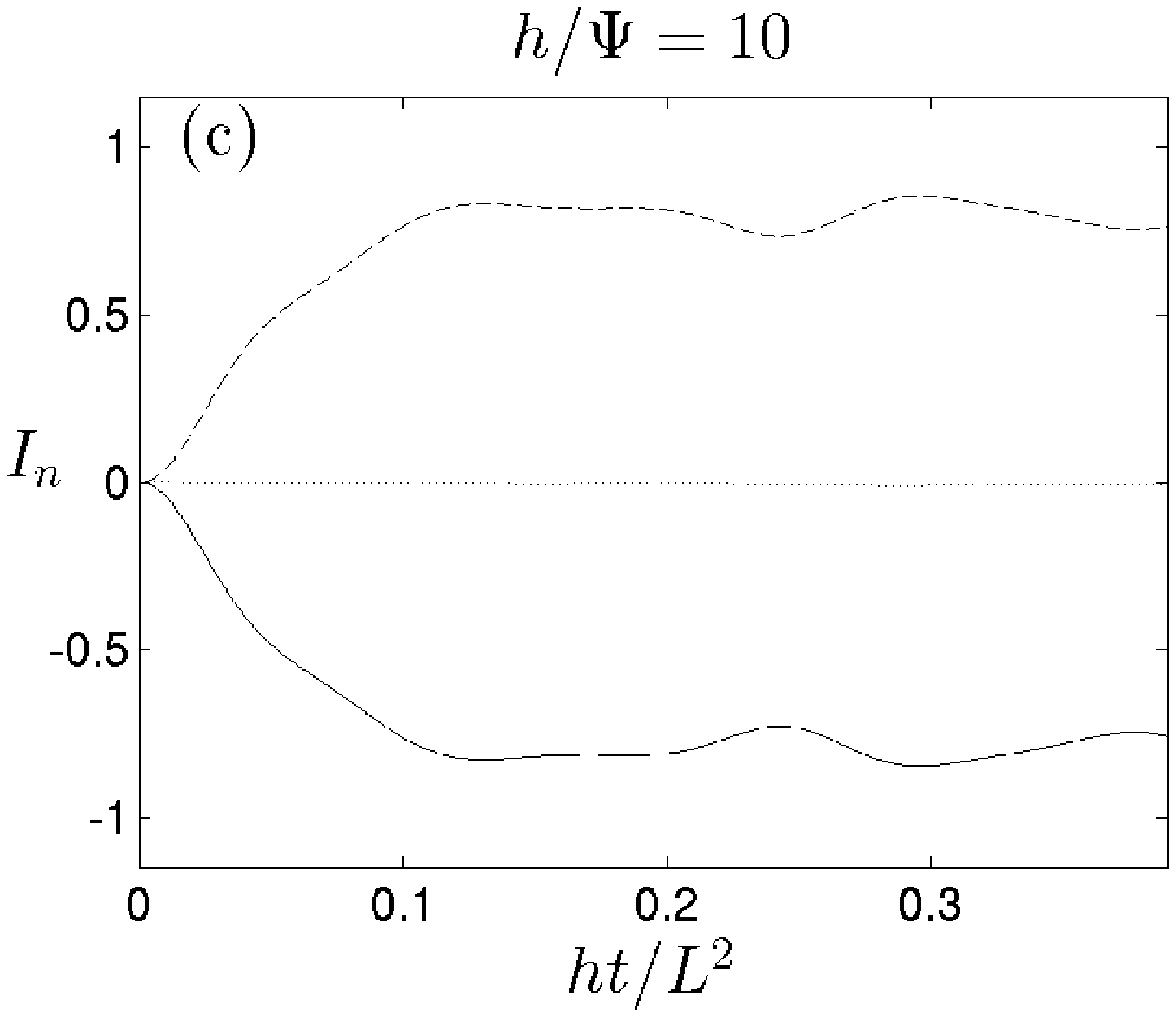}
\end{tabular}
\caption{Evolution of the integrals $I_1$ (dotted line), $I_2$
  (dashed line) and $I_3$ (thin solid line) in
  (\ref{eq:PE_conservation}) for $h/\Psi=0.2$ (a), $1$ (b) and $10$
  (c). $I_1$, $I_2$ and $I_3$ are scaled by the area of the domain. The inset in (a) represents a zoom on very short times.}
 \label{fig:I1_I2_I3_corr}
\end{center}
\end{figure}

The behaviour of $|M|$ is reflected in the evolution of $I_1$,
$I_2$ and $I_3$ displayed in figure \ref{fig:I1_I2_I3_corr}.  The
generation of small scales by refraction subsequently modulated
by dispersion leads to an increase of $I_2$. This is balanced by
the decrease of $I_3$ at short times for all values of $h/\Psi$
and for all times for $h/\Psi=10$, as predicted by our
scaling argument. Conversely, for $h/\Psi=0.2$ we obtain $I_1 \simeq - I_2$,
indicative of the balance between advection and dispersion noted
above (in this case, the amplitude of these quantities decrease slightly in time because of numerical dissipation).  For $h/\Psi=1$, all terms $I_1$, $I_2$ and $I_3$ have
similar final amplitudes.  Note that the amplitude of $I_2$
is %
roughly 
the same in the three simulations.
From the scaling $|M-1|=O(\Psi/h)$ mentioned above, we infer that $I_2$ is
proportional to $\Psi^2/L^2$, and similarly for $I_3$.  This
justifies a posteriori the inclusion of the overall factor $h$ in
the definition of $I_1$, $I_2$ and $I_3$.

\begin{figure}
\begin{center}
\begin{tabular}{ccc}
\includegraphics[width=.3\textwidth]{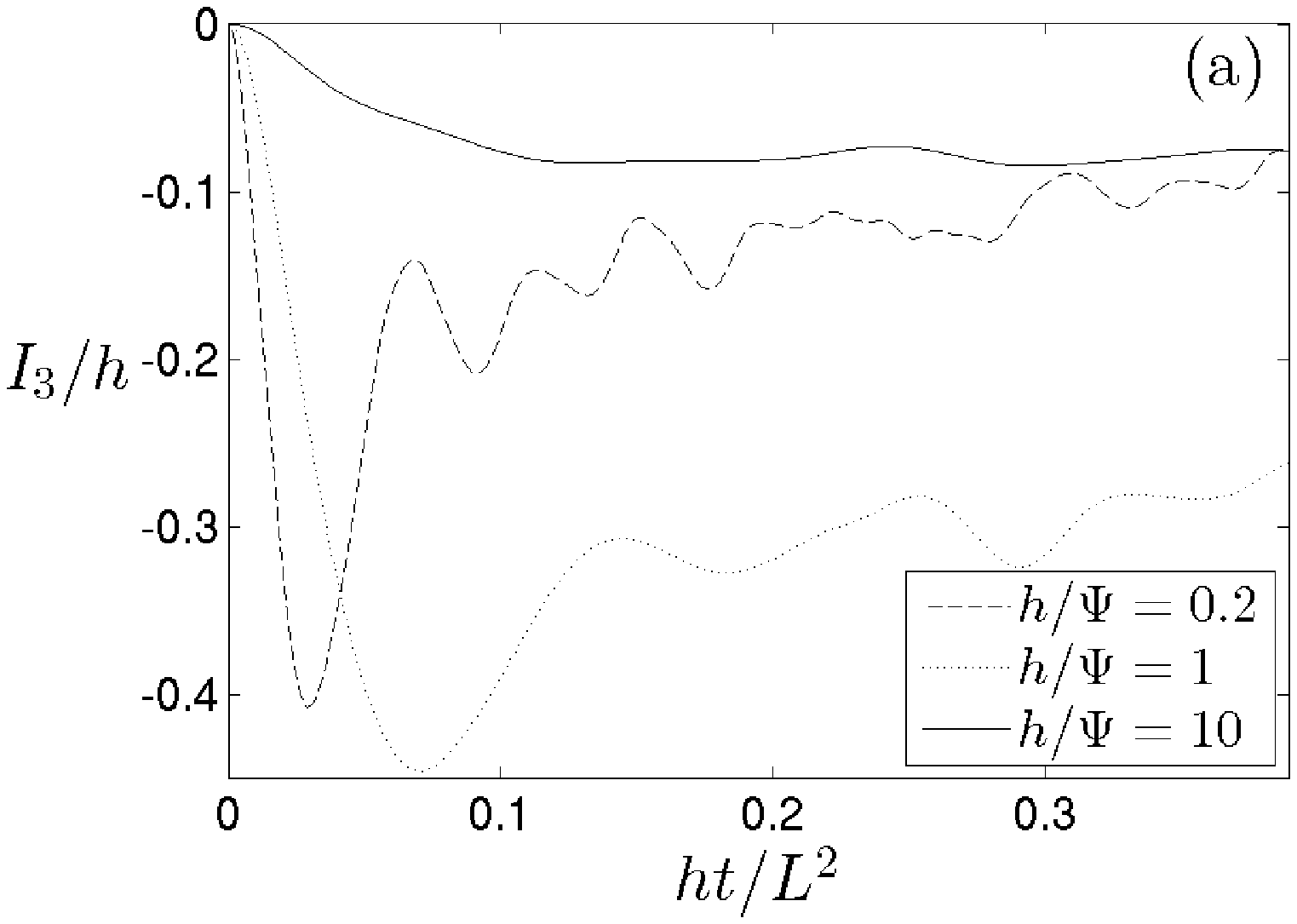}&
\includegraphics[width=.3\textwidth]{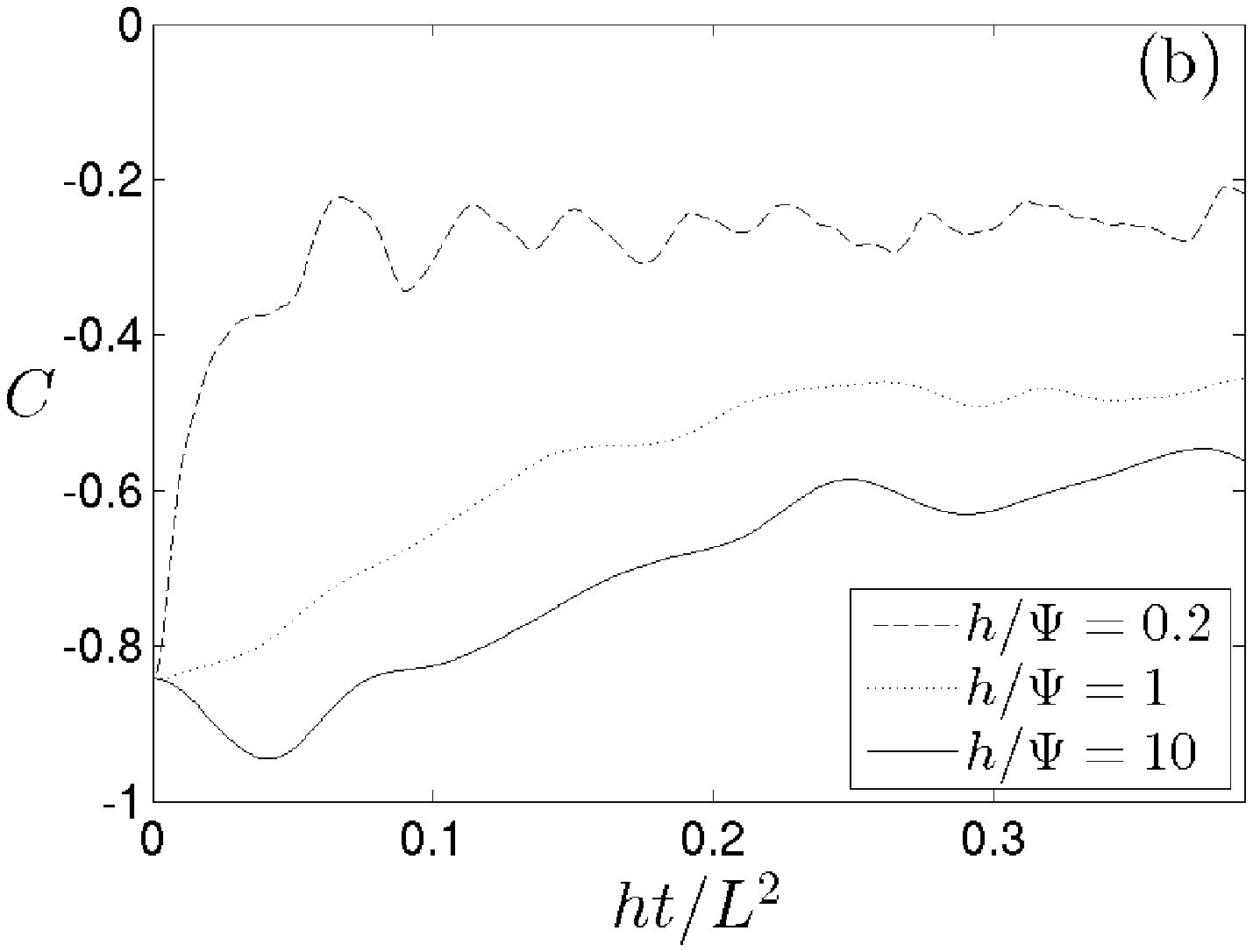}
\end{tabular}
\caption{(a) Spatial covariance $I_3/h$ and (b) correlation $C$ between $|M|^2$ and $\Delta \psi/2$ for $h/\Psi=0.2,\, 1$ and $10$.}
\label{fig:I3C}
\end{center}
\end{figure}

We now turn to the concentration of NIW energy in
anticyclones. This is best quantified by $I_3/h$, that is, the
covariance between $\Delta \psi/2$ and $|M|^2$. An alternative
diagnostic is the correlation between $|M|^2$ and $\Delta
\psi/2$, namely
\begin{equation}\label{eq:corr}
  C=\frac{I_3}{h  \left(\sigma_M^2 \iint{(\Delta \psi/2)^2}dxdy\right)^{1/2}}, \quad 
\textrm{where} \ \ 
\sigma_M^2 = \iint(|M|^4-1)dxdy
\end{equation}
is the variance of the NIW energy. We emphasise that $C$ focuses
on the match between the spatial patterns of NIW energy and
vorticity irrespective of their amplitudes; the more intuitive
colocation of high values of NIW energy with regions of
anticyclonic vorticity is measured by $I_3/h$. The evolution of
both $I_3/h$ and $C$ is shown for the three values of $h/\Psi$
in figure \ref{fig:I3C}. At the end of the simulations, the
covariance $I_3/h$ is largest for intermediate values of
$h/\Psi$. Small values of $h/\Psi$, lead to a strongly
heterogeneous $|M|^2$ (large $\sigma_M^2$, as seen on figure
\ref{fig:vort_M_field}g), but to relatively weak correlation of
$|M|^2$ and $\Delta \psi$ (figure \ref{fig:I3C}b, dashed line);
conversely, large values of $h/\Psi$ lead to a strong correlation
(figure \ref{fig:I3C}b, solid line) but weak heterogeneity of
$|M|^2$. The intermediate case displays both a relatively strong
$\sigma_M^2$ (figure \ref{fig:vort_M_field}h) and a strong
correlation (figure \ref{fig:I3C}b, dotted line), giving the
maximum covariance.

\section{Discussion} \label{sec:conclusion}
The results described in this paper in the context of the
reduced-gravity shallow-water system demonstrate that the
development of spatial heterogeneity in a homogeneous field of
NIWs due to the interaction with the barotropic vorticity field
is inevitably accompanied by a concentration of NIW-energy in
anticyclones. We emphasise that this result does not make
assumptions about the relative importance of dispersion (such as
the strong-dispersion approximation of \cite{YBJ} and
\cite{Balmforth98}) or the nature of the vorticity field and
smallness of advection \citep{Klein04}. Instead, it arises as a
consequence of a conservation law associated with \cite{YBJ}'s
model. Therefore, the concentration of NIW-energy in anticyclones
is a much more robust phenomenon than previously thought.
In particular, it is the strongest for the intermediate regime $h/\Psi=O(1)$, when refraction, dispersion and advection are all significant.

We note that (\ref{eq:4}) 
leads to an interesting conclusion for $h/\Psi \gg 1$.  In this
limit, $M \to 1$ as explained above so that the refraction term
in (\ref{eq:YBJ}) becomes approximately $i\Delta\psi/2$. By
(\ref{eq:4}) the time-derivative term has the same amplitude as
the refraction term in (\ref{eq:YBJ}), in contrast with previous
treatments of the strong-dispersion limit which neglect it
\citep{YBJ,Balmforth98}.  In this case, a more complete solution
of (\ref{eq:YBJ}) at short times is
\begin{equation}\label{eq:strong_disp}
M=1+\frac{1}{h}\iint\hat{\psi}(\mathbf{k})e^{i\mathbf{k}\cdot\mathbf{x}}(1-e^{-ih|\mathbf{k}|^2t/2}) d\mathbf{k}+O((\Psi/h)^{^2}),
\end{equation}
where $\hat{\psi}(\mathbf{k})$ is the Fourier transform of the
streamfunction $\psi$ at wave
number $\mathbf{k}$.  From (\ref{eq:strong_disp}), the adjustment
of the initial condition $M=1$ towards the `balanced state' with
$M=1+ \psi/h + O((\Psi/h)^{2})$ is accompanied by the emission of
$O(\Psi/h)$-amplitude waves. Although the latter part is missing
from \citet{YBJ} and \citet{Balmforth98}, (\ref{eq:strong_disp})
shows the correlation between $|M|$ and $\psi$ (hence the
anti-correlation between $|M|$ and the vorticity field) remains
true on average.  Note that a slow modulation of (\ref{eq:strong_disp}) should be added to describe the long-time behaviour of $M$ \citep{YBJ}.

In this paper, we make the strong assumption of a steady background flow. The advective time scale $L^2/\Psi$ typical of the flow evolution can be compared with the $O(L^2/h)$ time taken for $I_1$, $I_2$ and $I_3$ to reach saturation (see figure \ref{fig:I1_I2_I3_corr}) to conclude that the assumption of steadiness can be relaxed
when $h\gg\Psi$. When $h/\Psi=1$, saturation occurs for $ht/L^2\simeq 0.3$ so the impact of unsteadiness can be expected to remain weak. In the strong advection regime $h/\Psi \ll 1$, however, the time dependence of $\psi$ cannot be neglected and only the short-time solution described in \ref{sec:asymptotics} remains strictly valid.

Last, the assumption that NIWs initially have much larger horizontal
scales than the geostrophic flow is crucial for the results
reported here. Although this holds in many parts of the ocean,
NIWs can also be generated at scales similar to those of the
geostrophic flow (e.g., by a moving hurricane). The study of the
propagation of NIWs in a geostrophic flow with similar scales is
the subject of a forthcoming paper.

\medskip

\noindent
\textbf{Acknowledgements.} This research is funded by the UK
Natural Environment Research Council (grant NE/J022012/1).

\appendix 
\section{Reduced-gravity shallow-water YBJ model}\label{sec:derivation}
We begin by considering the reduced-gravity shallow-water system
linearized about a barotropic geostrophic flow
$(U,V)=(-\psi_y,\psi_x)$ (e.g., \citet{Klein04}). That is, we
assume $u,v\ll U,V$ and $\eta\ll H$, where $H$ is the
horizontally averaged depth of the top layer. We emphasize that,
because the geostrophic flow is barotropic (i.e.\ the same in the
top layer and beneath), there is no associated interface
slope. Under these assumptions, the NIW-velocity and layer-depth
perturbations obey
\begin{align}
\partial_t u+U\partial_x u+V\partial_y u+u\partial_x U+v\partial_y U -fv&=-g'\partial_x \eta,\label{eq:u}\\
\partial_t v+U\partial_x v+V\partial_y v+u\partial_x V+v\partial_y V +fu&=-g'\partial_y \eta,\label{eq:v}\\
\partial_t \eta+U\partial_x \eta+V\partial_y \eta+H(\partial_x u+\partial_y v)&=0.\label{eq:h}
\end{align}
Nondimensionalising using $(x,y)=L(x',y')$, $\psi=\Psi\psi'$,
$(u,v)=U_w(u',v')$, $\eta=HU_w/(fL)\,\eta'$ and $t=t'/f$, and
introducing a slow time-scale $\tau=\eps t$, we obtain the
following equation for the complex velocity $\mathcal{U}=u'+iv'$
by forming (\ref{eq:u})+i(\ref{eq:v}):
\begin{equation}\label{eq:complex_U}
\partial_t\mathcal{U}+i\mathcal{U}=-\eps \left(\partial_\tau\mathcal{U}+J(\psi,\mathcal{U})+2\eta_{\xi^*}+i\frac{\Delta\psi}{2}\mathcal{U}+2i\psi_{\xi^*\xi^*}\mathcal{U}^*\right),
\end{equation}
where $\xi=x+iy$ and $\Delta$ is the horizontal Laplacian. Primes in (\ref{eq:complex_U}) have been omitted for simplicity. We have assumed $\eps=\Psi/(fL^2)\sim g'H/(f^2L^2)\ll 1$; this corresponds to assuming a low Rossby number for the background flow and a small Burger number for the waves (i.e. waves oscillating at a frequency close to $f$).
The nondimensional version of (\ref{eq:h}) is 
\begin{equation}\label{eq:nondim_H}
\partial_t\eta+\eps J(\psi,\eta)+(\mathcal{U}_\xi+\mathcal{U}^*_{\xi^*})=0.
\end{equation}

An approximate solution can be sought by expanding $\mathcal{U}$ in powers of $\eps$: 
\begin{equation}\label{eq:expansion}
\mathcal{U}=\mathcal{U}^{(0)}+\eps\mathcal{U}^{(1)}+O(\eps^2).
\end{equation}
The leading order solution can be written as 
\begin{equation}\label{eq:U0}
\mathcal{U}^{(0)}=M(x,y,\tau)e^{-it},
\end{equation}
where $M$ describes the spatial and long-time modulation of the NIW-field. Inserting this form in (\ref{eq:nondim_H}) gives the leading order depth 
\begin{equation}\label{eq:h_solution}
\eta=-iM_\xi e^{-it}+\text{c.c.},
\end{equation}
where c.c. denotes complex conjugate.
The evolution equation for $M$ is found at the next order by
eliminating resonant terms and the dimensional version of the
resulting equation is 
then (\ref{eq:YBJ}). 
Eq.\ (\ref{eq:YBJ}) is also found for
continuously stratified flows when the geostrophic flow is
barotropic. In this case, it applies to the projection of the NIW
amplitude onto a single vertical mode, with $g'H/f$ replaced by
$fr_d^2$, where $r_d$ is the deformation radius of the vertical
mode. More details on the derivation (in the continuous
stratification case) can be found in \citet{YBJ}.  Note that
Eq.\ (\ref{eq:YBJ}) differs from that obtained by
\citet{Falkovich1994} and \citet{Reznik2001} for the
shallow-water model. This is because they consider a single-layer
model in which the geostrophic flow is balanced by a sloping free
surface. Our assumption of barotropic geostrophic flow and
consequent absence of interface slope is more relevant to the
oceanic context where geostrophic flows typically have vertical
scales much larger than the mixed-layer depth.

\section{NIW energy in the absence of a background flow}\label{sec:energy}
The energy associated with the linearized reduced-gravity shallow-water system (\ref{eq:u})--(\ref{eq:h}) in the absence of a flow ($U=V=0$) is 
\begin{equation}\label{eq:Energy_SW_lin_adim}
E=\iint\frac{1}{2}(u^2+v^2+\eps \eta^2)dxdy,
\end{equation}
using the non-dimensionalisation of Appendix \ref{sec:derivation}.
As expected, (\ref{eq:Energy_SW_lin_adim}) indicates that NIWs have much more kinetic than potential energy.
Inserting expansion (\ref{eq:expansion}) and solutions
(\ref{eq:U0}) and (\ref{eq:h_solution}) into
(\ref{eq:Energy_SW_lin_adim}) gives $E=E_0+\eps E_1+O(\eps^2)$, 
where 
$$E_0=\iint\frac{1}{2}|M|^2dxdy \ \ \textrm{and} \ \
E_1=\iint \Big(2|M_\xi|^2+Me^{-it}\mathcal{U}_1^*-M_\xi^2e^{-2it}+\text{c.c.}\Big)dxdy.$$
$E_0$ is the NIW-kinetic energy appearing in
(\ref{eq:KE_conservation}). Because $\mathcal{U}_1$ varies as
$e^{it}$ (since secular terms were removed, see \citet{YBJ},
equation (2.25)), the fast-time average of $E_1$ is 
just the first term, 
which is clearly proportional 
to $I_2$.

Thus, for $\psi=0$, conservation of total energy averaged over
fast-time gives (\ref{eq:KE_conservation}) at leading order and
(\ref{eq:PE_conservation}) at the next order.  Note that
(\ref{eq:KE_conservation})--(\ref{eq:PE_conservation}) are exact
conservation laws for the YBJ model but only adiabatic
invariants, i.e.\ approximate conservation laws, for the parent
shallow-water model \citep[e.g.,][]{Cotter-Reich}.

\bibliographystyle{jfm}
\bibliography{./biblio}

\begin{thebibliography}{18}
\expandafter\ifx\csname natexlab\endcsname\relax\def\natexlab#1{#1}\fi

\bibitem[Balmforth {\em et~al.\/}(1998)Balmforth, {Llewellyn Smith} \&
  Young]{Balmforth98}
{\sc Balmforth, N.J., {Llewellyn Smith}, S.G. \& Young, W.R.} 1998 Enhanced
  dispersion of near-inertial waves in an idealized geostrophic flow. {\em J.\
  Mar. \ Res.\/} {\bf 56}, 1--40.

\bibitem[Cotter \& Reich(2004)]{Cotter-Reich}
{\sc Cotter, C.~J. \& Reich, S.} 2004 Adiabatic invariance and applications:
  From molecular dynamics to numerical weather prediction. {\em BIT Numer.\
  Math.\/} {\bf 44}, 439--455.

\bibitem[Cushman-Roisin(1994)]{Cushman}
{\sc Cushman-Roisin, B.} 1994 {\em Introduction to Geophysical Fluid
  Dynamics\/}. Prentice Hall.

\bibitem[Danioux {\em et~al.\/}(2008)Danioux, Klein \& Rivi\`{e}re]{Danioux08a}
{\sc Danioux, E., Klein, P. \& Rivi\`{e}re, P.} 2008 Propagation of wind energy
  into the deep ocean through a fully turbulent mesoscale eddy field. {\em J.\
  Phys.\ Oceanogr.\/} {\bf 38}, 2224--2241.

\bibitem[Elipot {\em et~al.\/}(2010)Elipot, Lumpkin \& Prieto]{Elipot2010}
{\sc Elipot, S., Lumpkin, R. \& Prieto, G.} 2010 Modification of inertial
  oscillations by the mesoscale eddy field. {\em J.\ Geophys.\ Res.\/} {\bf
  115}, C09010.

\bibitem[Falkovich {\em et~al.\/}(1994)Falkovich, Kuznetsov \&
  Medvedev]{Falkovich1994}
{\sc Falkovich, G., Kuznetsov, E. \& Medvedev, S.} 1994 Nonlinear interaction
  between long inertio-gravity waves and rossby waves. {\em Nonlin. Processes
  in Geophys.\/} {\bf 1}, 168--171.

\bibitem[Ferrari \& Wunsch(2009)]{Ferrari2009}
{\sc Ferrari, R. \& Wunsch, C.} 2009 Ocean circulation kinetic energy:
  Reservoirs, sources, and sinks. {\em Annu.\ Rev.\ Fluid\ Mech.\/} {\bf 41},
  253--282.

\bibitem[Granata {\em et~al.\/}(1995)Granata, Wiggert \& Dickey]{Granata95}
{\sc Granata, T., Wiggert, J. \& Dickey, T.} 1995 Trapped, near-inertial waves
  and enhanced chlorophyll distributions. {\em J.\ Geophys.\ Res.\/} {\bf 100
  (C10)}, 20793--20804.

\bibitem[Joyce {\em et~al.\/}(2013)Joyce, Toole, Klein \& Thomas]{Joyce2013}
{\sc Joyce, T.M., Toole, J.M., Klein, P. \& Thomas, L.N.} 2013 A near-inertial
  mode observed within a gulf stream warm-core ring. {\em J.\ Geophys.\ Res.\/}
  {\bf 118}, 1797--1806.

\bibitem[Klein {\em et~al.\/}(2004)Klein, {Llewellyn Smith} \&
  Lapeyre]{Klein04}
{\sc Klein, P., {Llewellyn Smith}, S. \& Lapeyre, G.} 2004 Organization of
  near-inertial energy by an eddy field. {\em Q.\ J.\ R.\ Meteorol.\ Soc.\/}
  {\bf 130}, 1153--1166.

\bibitem[Kunze(1985)]{Kunze85}
{\sc Kunze, E.} 1985 Near-inertial wave propagation in geostrophic shear. {\em
  J.\ Phys.\ Oceanogr.\/} {\bf 15}, 544--565.

\bibitem[Kunze \& Sanford(1984)]{Kunze84}
{\sc Kunze, E. \& Sanford, T.B.} 1984 Observations of near-inertial waves in a
  front. {\em J.\ Phys.\ Oceanogr.\/} {\bf 14}, 566--581.

\bibitem[Lee \& Niiler(1998)]{Lee98}
{\sc Lee, D.-K. \& Niiler, P.~P.} 1998 The inertial chimney: the near-inertial
  energy drainage from the ocean surface to the deep layer. {\em J.\ Geophys.\
  Res.\/} {\bf 103 (C4)}, 7579--7591.

\bibitem[{Llewellyn Smith}(1999)]{Llewellyn99}
{\sc {Llewellyn Smith}, S.} 1999 Near-inertial oscillations of a barotropic
  vortex: Trapped modes and time evolution. {\em J.\ Phys.\ Oceanogr.\/} {\bf
  29}, 747--761.

\bibitem[Reznik {\em et~al.\/}(2001)Reznik, Zeitlin \& {Ben
  Jelloul}]{Reznik2001}
{\sc Reznik, G.M., Zeitlin, V. \& {Ben Jelloul}, M.} 2001 Nonlinear theory of
  geostrophic adjustment. part 1. rotating shallow-water model. {\em J.\ Fluid\
  Mech.\/} {\bf 445}, 93--120.

\bibitem[Xie \& Vanneste(2015)]{Xie2015}
{\sc Xie, J.-H. \& Vanneste, J.} 2015 A generalised-lagrangian-mean model of
  the interactions between near-inertial waves and mean flow. {\em J.\ Fluid\
  Mech.\/} \itshape In revision.

\bibitem[Young \& {Ben Jelloul}(1997)]{YBJ}
{\sc Young, W.R. \& {Ben Jelloul}, M.} 1997 Propagation of near-inertial
  oscillations through a geostrophic flow. {\em J.\ Mar. \ Res.\/} {\bf 55},
  735--766.

\bibitem[Zhai {\em et~al.\/}(2005)Zhai, Greatbach \& Zhao]{Zhai05}
{\sc Zhai, X., Greatbach, R.~J. \& Zhao, J.} 2005 Enhanced vertical propagation
  of storm-induced near-inertial energy in an eddying ocean channel model. {\em
  Geophys.\ Res.\ Lett.\/} {\bf 32}, L18602.

\end{thebibliography}

\end{document}